\documentclass{moriond}

\usepackage{amssymb}
\usepackage{amsmath}
\usepackage{xspace}

\newcommand{\sqrtsnn}{\sqrt{\rm s_{_{\mathrm{NN}}}}}
\newcommand{\sqrtsgaga}{\sqrt{\rm s_{_{\gamma\gamma}}}}

\newcommand{\gaga}{\gamma\gamma}

\providecommand{\bbbar}{b\overline{b}}
\providecommand{\ccbar}{c\overline{c}}

\providecommand{\qqbar}{q\overline{q}}

\newcommand{\mcfm}{{\sc mcfm}}
\newcommand{\madgraph}{{\sc madgraph}}

\newcommand*{\elm}{e.m.\@\xspace}

\newcommand{\cm}{c.m.\@\xspace}
\newcommand{\LInt}{\mathcal{L}_{\rm \tiny{int}}}

\def\cO#1{{{\cal{O}}}\left(#1\right)}

\newcommand{\Photo}{\includegraphics[height=30mm]{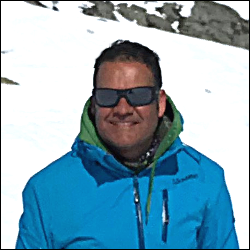}}

\begin{document}
\title{Higgs boson production in partonic and electromagnetic\\ interactions with heavy ions}


\author{David d'Enterria}
\address{CERN, EP Department, CH-1211 Geneva 23, Switzerland\vspace{0.2cm}}
\maketitle
\abstracts{
Higgs boson production in partonic and electromagnetic (photon-photon) interactions of light- and heavy-ions (A-A) 
at the LHC and future colliders is summarized. Parton-induced cross sections --including gluon-gluon, 
vector-boson fusion, and associated W,\,Z,\,t processes-- are computed at NNLO with \mcfm~8.0 using 
nuclear parton distribution functions. Photon-photon cross sections are computed with \madgraph~5.0
for ultraperipheral A-A interactions with both ions surviving the collision. In the center-of-mass energy 
range $\sqrtsnn \approx 5$--100~TeV, the ratio of electromagnetic-to-partonic Higgs 
cross sections is $\rm R_{e.m./parton}\approx 10^{-5}$--10$^{-4}$. At FCC energies, this ratio
is 10--100 times larger in A-A than in p-p thanks to the strong nuclear coherent $\gamma$ fluxes. 
The feasibility of Higgs boson measurements at LHC, HE-LHC, and FCC, in the most-favourable $\rm H\to\gaga,\bbbar$ 
decay channels in partonic and \elm\ interactions respectively, is determined taking into account standard 
acceptance and efficiency losses, plus selection criteria to minimize the respective 
continuum backgrounds. Whereas $3\sigma$ evidence for partonic and \elm\ Higgs production requires about 
$\times$35 and $\times$200 ($\times$7 and $\times$30) larger integrated luminosities than those expected 
for a nominal 1-month run at the LHC (HE-LHC), a $5\sigma$ observation of both production modes is warranted in 
just one FCC month.
}

\section{Introduction}
\label{sec:intro}

The Higgs boson remains unobserved in heavy-ion collisions at the LHC due to the lower nucleon-nucleon \cm\ energies ($\sqrtsnn$), 
as well as integrated luminosities ($\LInt$), in A-A compared to p-p collisions. 
Three planned or proposed CERN future hadron colliders: the high-luminosity LHC (HL-LHC)~\cite{HL_LHC_HE_LHC_AA}, 
high-energy LHC (HE-LHC)~\cite{HL_LHC_HE_LHC_AA,HE_LHC}, and Future Circular Collider (FCC)~\cite{FCC_AA},
will noticeably increase the available \cm\ energies and $\LInt$ values, thereby making the observation of the scalar boson potentially 
feasible in nuclear collisions. The motivations for such a measurement are twofold. First, when produced in standard 
parton-induced processes, the H boson can be used to probe the properties of the quark-gluon plasma produced in 
A-A collisions~\cite{dEnterria:2018bqi} (although this possibility has been challenged in Ref.~\cite{Ghiglieri:2019lzz}). 
Secondly, the Higgs boson can also be exclusively produced via A-A$\xrightarrow{\gamma\gamma}$(A)H(A) 
in ultraperipheral heavy-ion collisions (UPCs) where the ions survive their electromagnetic (\elm) interaction. 
This latter unique production mode exploits the huge \elm fields generated by the collective action of $Z$ individual 
proton charges, leading to photon-photon cross sections enhanced by 
up to factors $Z^4 \approx 5\cdot 10^{7}$, for Pb-Pb, compared to proton or e$^\pm$ beams~\cite{Baltz:2007kq}. 
The UPC production provides an independent measurement of the H-$\gamma$ coupling not based on Higgs decays 
but on $s$-channel production, and of the total Higgs width by combining it with the LHC diphoton decay 
branching ratio, via $\rm\Gamma_{\rm H}^{\rm tot}=\Gamma(H\to\gaga)/\mathcal{B}(H\to\gaga)$~\cite{Borden:1993cw}.

The parton-initiated cross sections --including gluon-gluon, vector-boson fusion, and associated 
W,\,Z,\,t processes-- have been studied for A-A collisions at the LHC and FCC in Ref.~\cite{dEnterria:2017jyt}, including their expected yields 
after analysis cuts in the cleanest $\rm H\to\gaga$ decay. The Higgs UPC-production 
was first considered 30 years ago~\cite{higgs_upc}, and detailed studies of the dominant ${\rm H}\to\bbbar$ decay channel, 
including realistic acceptance and efficiencies for signal and $\gaga\to\bbbar$ continuum background presented 
in Refs.~\cite{dEnterria:2009cwl} for LHC, HE-LHC, and FCC. This work here extends previous studies by
including also light-ion collisions, and compares the results of both production modes.

\begin{figure*}[htpb!]
\centering
\includegraphics[width=0.47\textwidth]{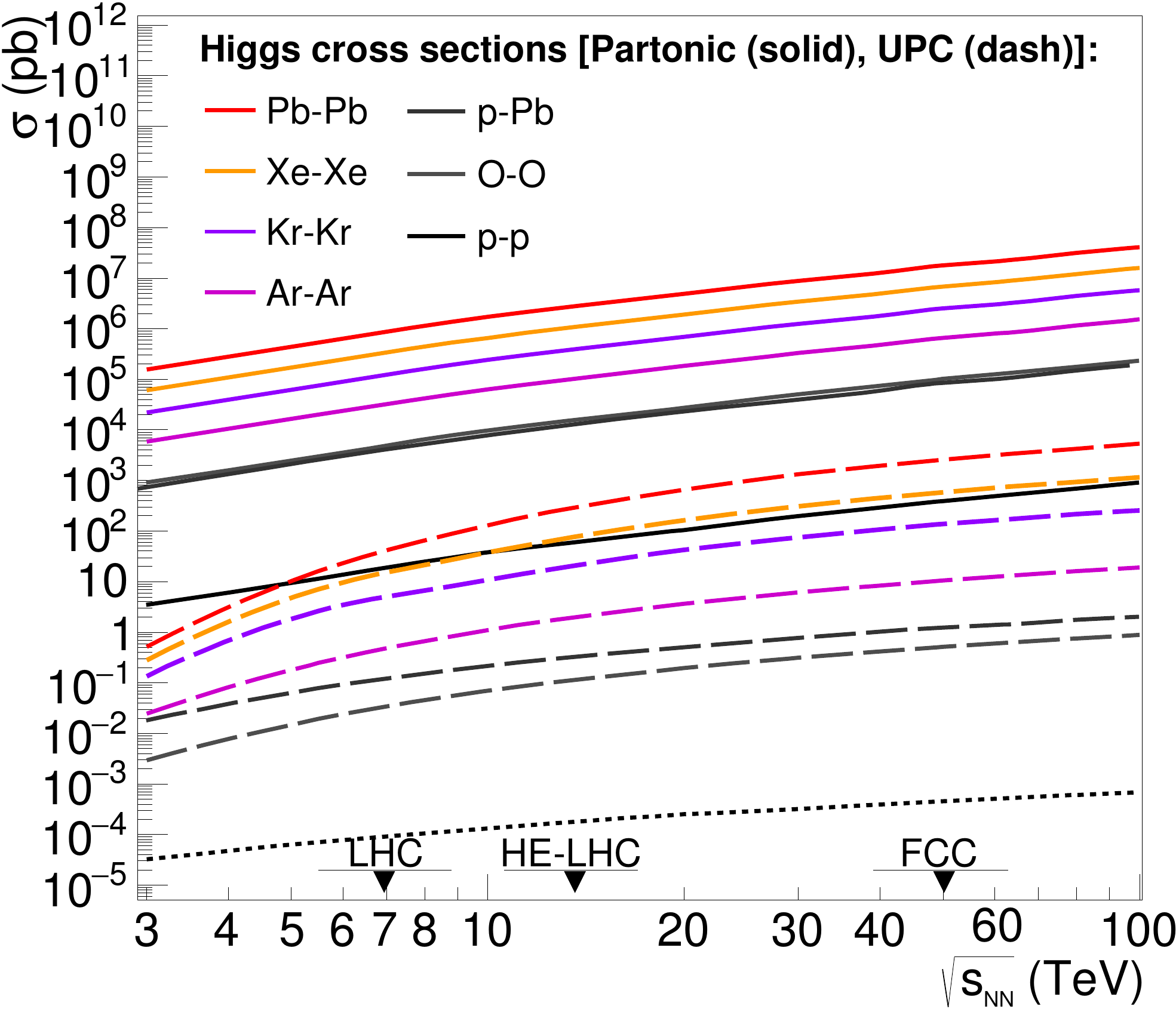}\hspace{0.2cm}
\includegraphics[width=0.45\textwidth]{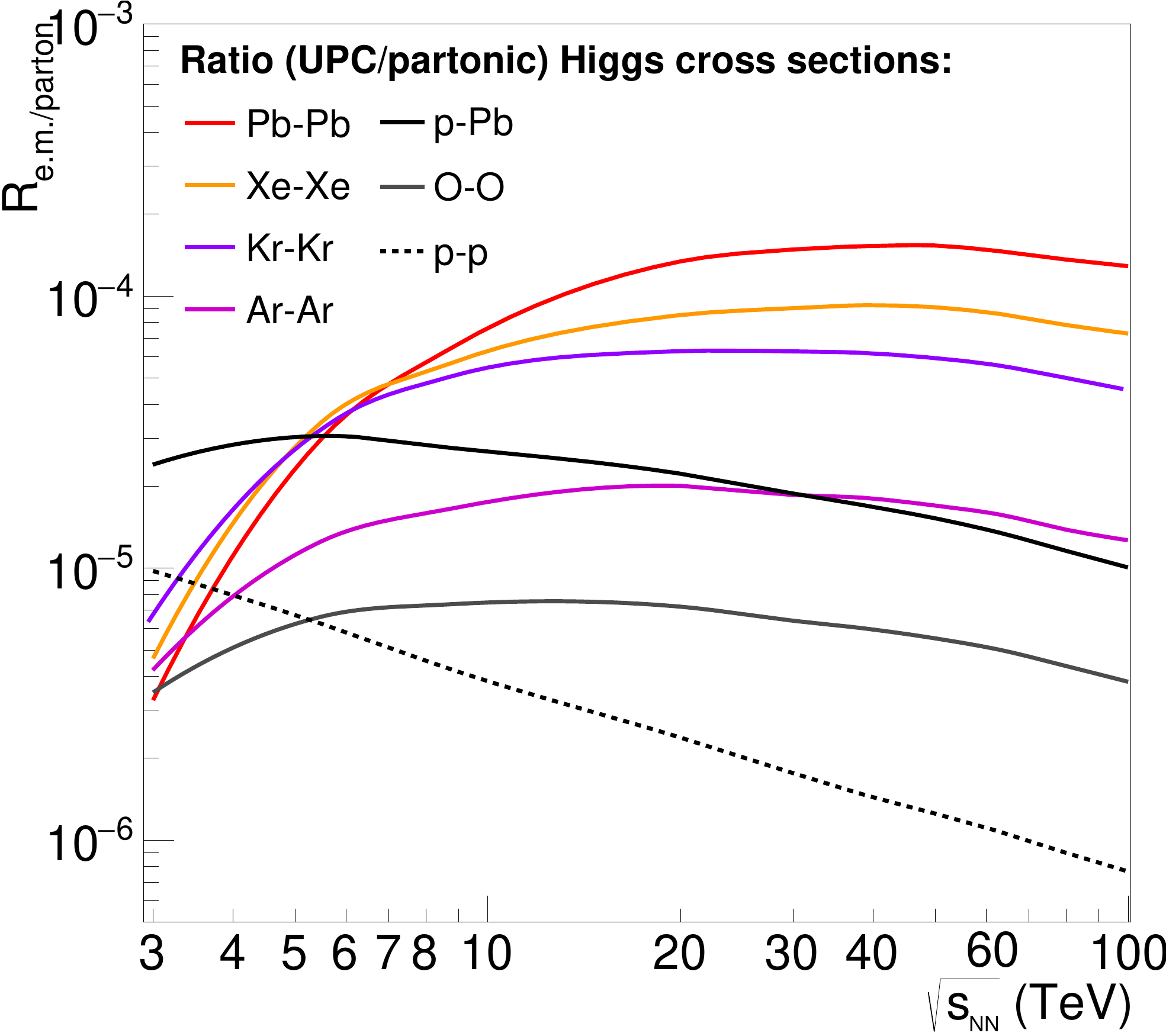}
\caption{Partonic (solid) and \elm\ (dashed) Higgs cross sections  (left) and their ratio (right) 
in Pb-Pb, Xe-Xe, Kr-Kr, Ar-Ar, O-O, p-Pb, and p-p collisions as a function of \cm\ energy over $\sqrtsnn = 3$--100~TeV.} 
\label{fig:xsec}
\end{figure*}

\section{Higgs boson in partonic A-A collisions}
\label{sec:}

The Higgs cross sections in p-Pb and Pb-Pb collisions over $\sqrtsnn \approx 3$--100~TeV 
have been computed at NNLO accuracy in Ref.~\cite{dEnterria:2017jyt} with \mcfm\ (v.8.0)~\cite{Boughezal:2016wmq}, 
using the CT10 proton PDFs~\cite{Lai:2010vv} and EPS09 nuclear PDFs (nPDFs, including its 30 eigenvalues sets)~\cite{Eskola:2009uj},
and fixing the default renormalization and 
factorization scales at $\rm \mu_R=\mu_F=m_{\rm H}/2$. The PDF uncertainties, obtained adding in quadrature the CT10 and EPS09 
uncertainties, are $\sim$10\%. The computed nucleon-nucleon cross sections are scaled by the mass number A or
A$^2$ to obtain the corresponding p-A or A-A cross sections respectively, including those for 
future runs with lighter-ions (Xe-Xe, Kr-Kr, Ar-Ar, and O-O), not considered so far.
Compared to the corresponding p-p results at a given \cm\ energy, antishadowing nPDF modifications 
slightly enhance (deplete) the nucleon-nucleon yields by $\sim$3\% at the LHC (FCC).
The $\cO{5-10\%}$ theoretical $\mu_{R,F}$ scale uncertainties 
cancel out in the ratios of (p-Pb,Pb-Pb)/(p-p) cross sections at the same colliding energy.
The collision energy dependence of the parton-induced Higgs cross sections for all colliding systems
is shown Fig.~\ref{fig:xsec} (left, solid curves).
The cross sections increase by about a factor of $\times$4 and $\times$20 between LHC 
($\sigma_{\rm PbPb\to H+X}\approx$~500~nb), and HE-LHC ($\sim$2~$\mu$b) and FCC 
($\sim$11.5~$\mu$b) respectively.

\begin{figure*}[htpb!]
\centering
\includegraphics[width=0.45\textwidth,height=5.2cm]{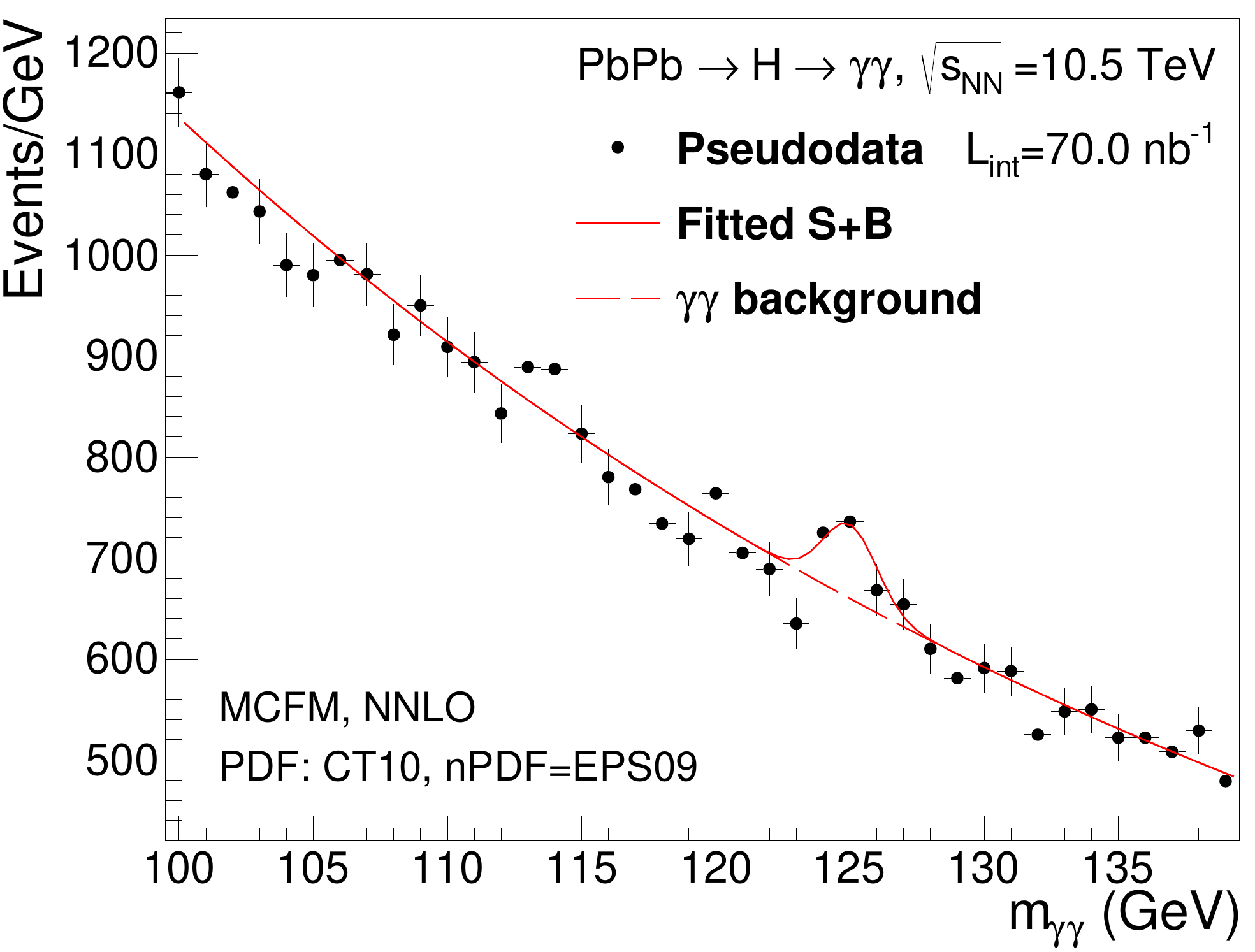}\hspace{0.5cm}
\includegraphics[width=0.45\textwidth,height=5.2cm]{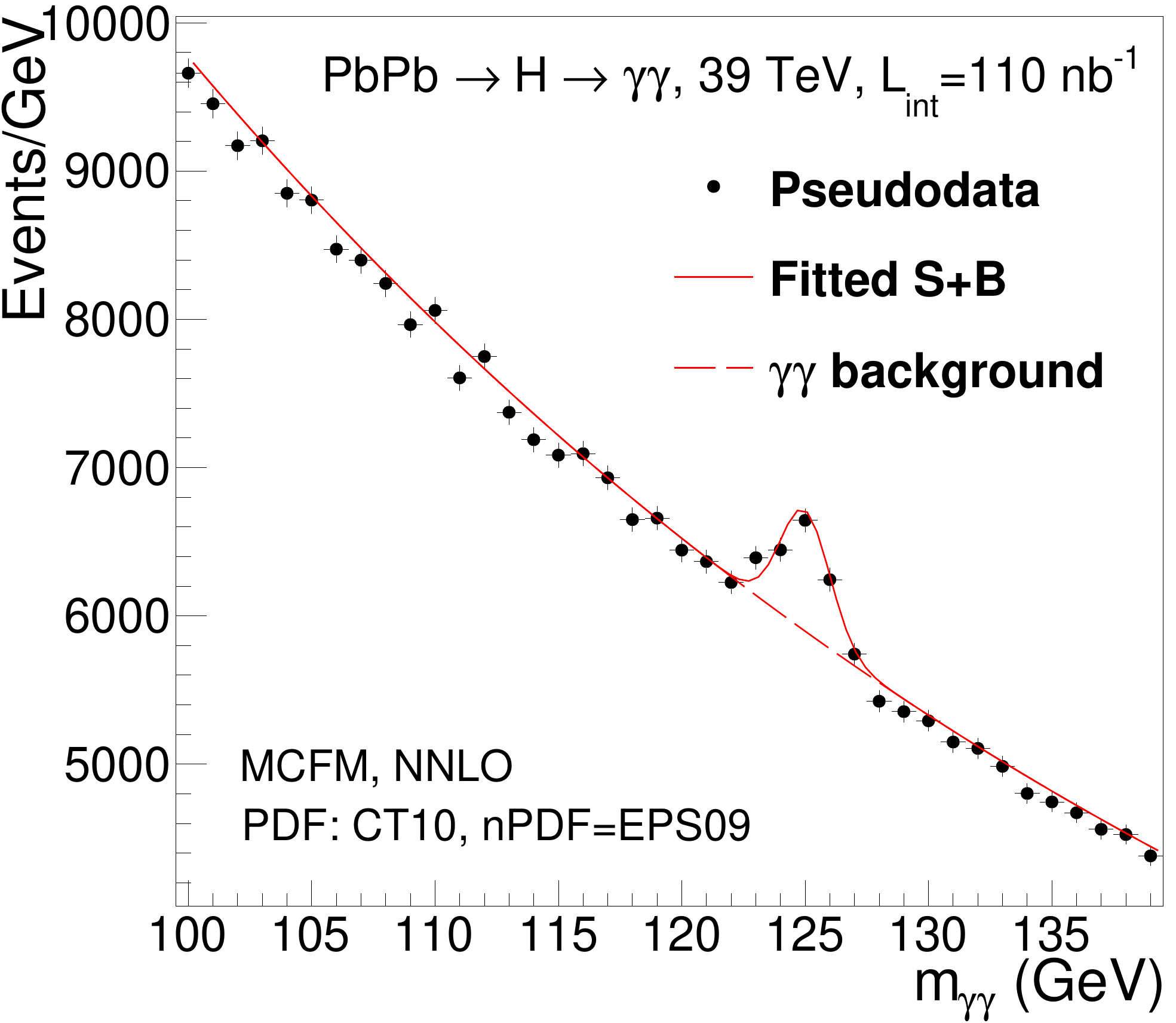}
\caption{Expected diphoton invariant mass distributions in Pb-Pb at 
HE-LHC ($\LInt$~=~70~nb$^{-1}$, left) and 
FCC (nominal 1-month $\LInt$~=~110~nb$^{-1}$, right) for a Higgs boson 
signal on top of the $\gaga$ backgrounds after cuts.} 
\label{fig:Higgs}
\end{figure*}

A first Higgs measurement in nuclear collisions will use the clean diphoton and 4-lepton final-state channels
as done in p-p~\cite{Higgs_LHC}, with very low branching ratios but small and/or controllable backgrounds.
After branching ratios ($\mathcal{B} = 0.23\%$ for $\gaga$, 0.012\% for $4\,\ell$), and taking into account 
typical ATLAS/CMS analysis cuts\footnote{$p_T^\gamma > 30,40$~GeV, $|\eta^\gamma| < 2.5,\,5$\,(LHC,\,FCC); 
$p_T^\ell > 20,15,10,10$~GeV, $|\eta^\ell| < 2.5,\,5$\,(LHC, FCC); $R_{\rm isol}^{\gamma,\ell} = 0.3$.},
which result in $\sim$50\% acceptance and efficiency signal losses, 
one expects about 10, 50, and 1\,500 Higgs bosons per month in Pb-Pb at the LHC (5.5 TeV), 
HE-LHC (10.6 TeV), and FCC (39 TeV) with $\LInt =$~10, 10, and 110~nb$^{-1}$ 
respectively, on top of the corresponding $\gaga$ and $4\ell$ non-resonant backgrounds. 
In the $\gaga$ case, the backgrounds include the irreducible QCD diphoton continuum plus
30\% of events coming from misidentified $\gamma$-jet and jet-jet processes.
For the nominal HL-LHC and HE-LHC luminosities, the significances of the diphoton signal 
(S) over the background (B), computed via S/$\sqrt{\rm B}$ at the Higgs peak,
are 0.5 and 1.2$\sigma$, and thus one would need a factor of $\times$35 and $\times$7 larger integrated 
luminosities to reach $3\sigma$ evidence at both machines. A straightforward way to gain 
a factor of $\sim$10 in $\LInt$ would be to run one full-year, instead of one heavy-ion month, 
but even in this case 
evidence at the HL-LHC would require an extra $\times$3 increase in the instantaneous luminosity. 
The situation is much more favourable at HE-LHC and FCC.
Figure~\ref{fig:Higgs} shows the expected diphoton invariant mass distributions for Pb-Pb 
at HE-LHC ($\times$7 the nominal 1-month luminosity, left) and FCC (nominal luminosity, right). 
The significances of the H$(\gaga)$ peaks are 3$\sigma$ and 5.5$\sigma$ respectively.

\section{Higgs boson in electromagnetic A-A collisions}
\label{sec:}

The \madgraph~5 (v.2.6.5) Monte Carlo event generator~\cite{madgraph} is employed to compute the UPC cross sections
for Higgs boson and diquark continuum ($\bbbar$, as well as misidentified $\ccbar$ and light-quark $\qqbar$, 
pairs)~\cite{dEnterria:2009cwl}, by convolving the Weizs\"acker-Williams equivalent $\gamma$ fluxes of the ions 
(as a function of impact parameter $b$)~\cite{Jackson} and/or protons~\cite{Budnev:1974de} with the elementary 
$\gaga \to \rm H$ cross section (with H-$\gamma$ coupling parametrized in the Higgs effective field theory). 
We exclude nuclear overlap by imposing $b_1 > R_{A_{1}}$ and $b_2 > R_{A_{2}}$ for each $\gamma$ flux (where
$R_{A}$ is the radius of nucleus A), and applying a correcting factor~\cite{Cahn:1990jk} on the final cross section 
that depends on the ratio of Higgs mass over $\sqrtsgaga$. The computed exclusive $\gaga\to$\,Higgs cross sections 
are shown in Fig.~\ref{fig:xsec} (left, dashed curves) as a function of $\sqrtsnn$ for all colliding systems. 
The larger the charge of the ions, the larger the UPC cross sections.
In the Pb-Pb case, the Higgs cross sections increase by about a factor of $\times$10 and $\times$100 between LHC 
($\sigma_{\rm \gaga\to\,H}\approx$~15~pb), and HE-LHC ($\sim$150~pb) and FCC ($\sim$1.8~nb) respectively.
Note that, below $\sqrtsnn\approx 5$~TeV, the \elm\ production cross section in Pb-Pb 
is larger than the {\it partonic} one in p-p collisions.
The ratio of \elm-to-partonic Higgs cross sections is shown in Fig.~\ref{fig:xsec} (right).
For A-A collisions, one finds $\rm R_{e.m./parton}\approx 10^{-5}$--10$^{-4}$ due to the strong nuclear 
coherent $\gamma$ fluxes. This ratio in practice amounts to $\rm R_{e.m./parton}\approx 0.01$--0.03 when 
one further accounts for the $\mathcal{B} = 0.27\%,56\%$ of the most-favourable $\rm H\to\gaga,\bbbar$ 
decay channels for a measurement in partonic and \elm\ interactions respectively. As a function of \cm\ 
energy, $\rm R_{e.m./parton}$ decreases for p-p (as the partonic cross sections increase faster than 
the \elm\ ones) but increases for A-A (as the energy of the coherent $\gamma$ fluxes rise) up to a point 
where it levels off and starts to drop. At FCC energies, the ratio is 10--100 larger in A-A than in p-p.

\begin{figure*}[htpb!]
\centering
\includegraphics[width=0.45\textwidth,height=4.9cm]{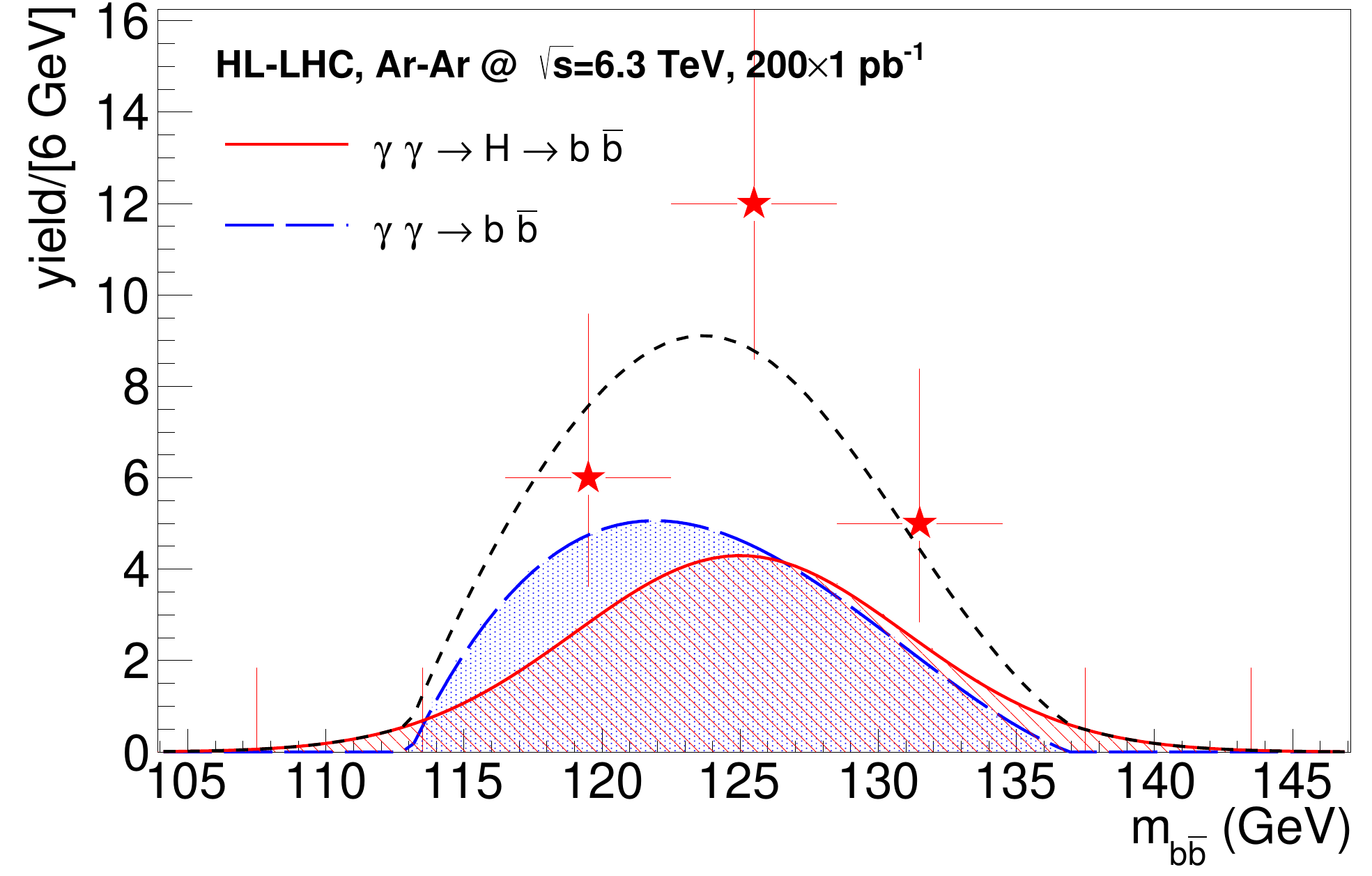}\hspace{0.5cm}
\includegraphics[width=0.45\textwidth,height=4.9cm]{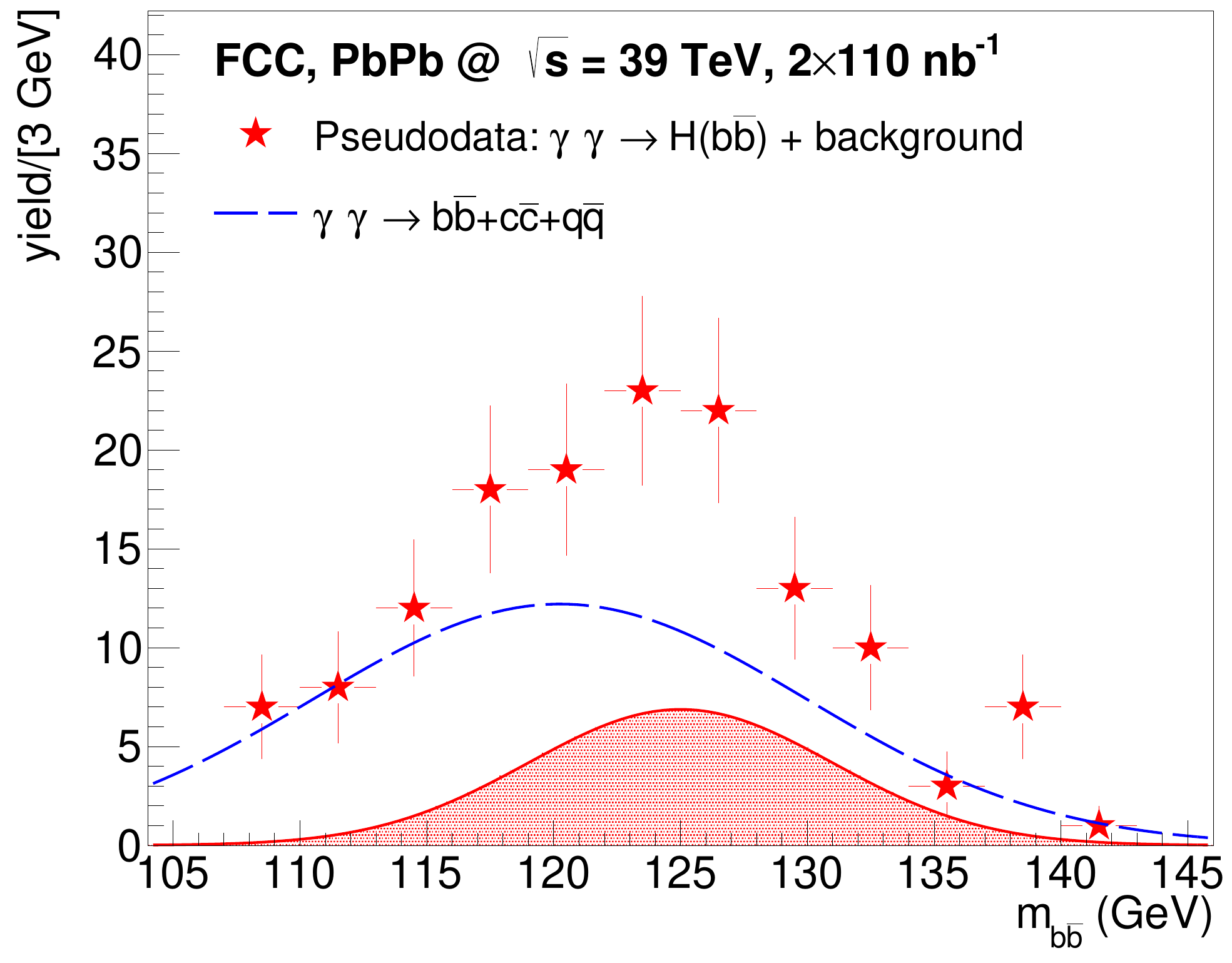}
\caption{Expected invariant mass distributions for b-jet pairs from $\gaga\to {\rm H}(\bbbar)$ signal 
(hatched red Gaussian) and $\bbbar+\ccbar+\qqbar$ continuum (blue curve) 
in ultraperipheral Ar-Ar (HL-LHC, left) and Pb-Pb (FCC, right) collisions~\protect\cite{dEnterria:2009cwl}.} 
\label{fig:Higgs_UPC}
\end{figure*}

The feasibility of measuring \elm\ exclusive Higgs production is studied in its dominant H\,$\to\bbbar$ decay. 
Taking into account the expected beam luminosities, the most competitive system to try a measurement 
is Ar-Ar (Kr-Kr) at the HL-LHC (HE-LHC).
At the FCC, the combination of $\times$100 larger cross sections and $\times10$ larger $\LInt$  
yield $\sim$200 Higgs produced per month ($\sim$110 in the $\bbbar$ channel). 
The following acceptance and reconstruction performances are assumed: jet reconstruction over $|\eta|<2$ 
($5$ for FCC), 7\% b-jet energy resolution (resulting in $\sim$6~GeV dijet mass resolution at the Higgs peak), 
70\% b-jet tagging efficiency, and 5\% (1.5\%) b-jet mistagging probability for a $c$ 
($q$) quark. These selection criteria lead to a $\sim$50\% signal loss, and a total reduction of the misidentified 
$\ccbar$ and $\qqbar$ non-resonant backgrounds by factors of $\sim$400 and $\sim$4500 respectively. 
The remaining irreducible $\bbbar$ continuum is further suppressed through kinematical cuts on the jet 
$p_{T}^{jet}\approx$~55--67~GeV, invariant dijet mass $m_{\bbbar}\approx$~116--134~GeV, and 
dijet angular distribution $|\cos \theta_{j_{1}j_{2}}| < 0.5$.
The final signal significance is derived from the number of counts around the Gaussian 
Higgs peak over the dijet continuum. Evidence for UPC Higgs at the HL-LHC requires $\times$200 times 
larger-than-nominal $\LInt$ values in Ar-Ar(6.3~TeV), so as to reach S/$\sqrt{\rm B}\approx 10/\sqrt{12}\approx 3$ 
(Fig.~\ref{fig:Higgs_UPC}, left). The same analysis for Kr-Kr(12.5~TeV) at the HE-LHC indicates that 
$\times$30 larger $\LInt$ values are needed than currently designed.
On the other hand, Pb-Pb collisions at $\sqrtsnn$~=~39 GeV with the FCC nominal 
$\LInt = 110\;\mbox{nb}^{-1}$ per month, yield $\sim$20 signal counts over about the same 
number for the sum of backgrounds. Reaching a $5\sigma$ statistical significance would just require 
to combine the measurements of the first run from two different experiments or accumulating two 
1-month runs in a single one (Fig.~\ref{fig:Higgs_UPC}, right). 

\vspace{-0.4cm}
\section{Conclusions}
\label{sec:}

The expected Higgs production cross sections and yields from parton-induced and exclusive photon-photon 
processes in heavy- and light-ion collisions have been studied in the diphoton and $\bbbar$ final 
states respectively at the HL-LHC, HE-LHC, and FCC.
Evidence for the scalar boson, over the expected continuum backgrounds, seems out of reach at the HL-LHC with 
the currently expected integrated luminosities, but can be possible in the partonic mode at HE-LHC. 
Full observation in both production modes requires a future hadron collider such as the FCC.



\end{document}